\documentstyle[aps,epsfig]{revtex}
\begin{document}

\title{Proton spin structure and intrinsic motion of the constituents
\footnote{Talk prepared for the conference DIS2004}}

\author{Petr Z\'{a}vada}

\address{Institute of Physics, Academy of Sciences of the Czech Republic,\\
Na Slovance 2, CZ-182 21 Prague 8\\
E-mail: zavada@fzu.cz\\
}

\maketitle

\begin{abstract}
The spin structure of the system of quasifree fermions having total 
angular momentum $J=1/2$ is studied in a consistently covariant approach. Within
this model the relations between the spin functions are obtained. Their
particular cases are the sum rules Wanzura - Wilczek, Efremov - Leader -
Teryaev, Burkhardt - Cottingham and also the expression for the Wanzura -
Wilczek twist 2 term $g_{2}^{WW}$. With the use of the proton valence quark
distributions as an input, the corresponding spin functions are obtained.
The resulting structure functions $g_{1}$ and $g_{2}$ are well compatible
with the experimental data. Comparison with the basic formulas following from 
the standard quark-parton model reveals the importance of the quark intrinsic motion 
inside the target for the correct evaluation of the spin structure functions.
\end{abstract}

\section{Introduction}
In this talk some results following from the covariant quark-parton model
(QPM) will be shortly discussed, details of the model can be found in Refs. %
\cite{zav4}, \cite{zav5}. In this version of QPM valence quarks are
considered as quasifree fermions on mass shell. Momenta distributions
describing the quark intrinsic motion have spherical symmetry corresponding
to the constraint $J=1/2$, which represents the total angular momentum -
generally consisting of spin and orbital parts. I shall mention the
following items:

1. What sum rules follow from this approach for the spin structure functions 
$g_{1}$ and $g_{2}$?

2. How can these structure functions be obtained from the valence quark
distributions $u_{V}$ and $d_{V}$ - if the $\ SU(6)$ symmetry is assumed?
The results are compared with the existing experimental data.

3. Why the first moment $\Gamma _{1}$ calculated in this approach can be
substantially less, than the corresponding moment calculated within the
standard, non covariant QPM, which is based on the infinite momentum frame?

Recently, this model was generalized to include also the transversity
distribution, for details I refer to \cite{tra}.

\section{Model}
The model is based on the set of distribution functions $G_{k,\lambda }(%
\frac{pP}{M})$, which measure probability to find a quark in the state:%
\[
u\left( p,\lambda {\bf n}\right) =\frac{1}{\sqrt{N}}\left( 
\begin{array}{c}
\phi _{\lambda {\bf n}} \\ 
\frac{{\bf p}{\bf \sigma }}{p_{0}+m}\phi _{\lambda {\bf n}}%
\end{array}%
\right) ;\qquad \frac{1}{2}{\bf n\sigma }\phi _{\lambda {\bf n}}=\lambda
\phi _{\lambda {\bf n}},\qquad \lambda =\pm \frac{1}{2}, 
\]%
where$\ {\bf n}$ coincides with the direction of the proton polarization $%
{\bf J}$. Correspondingly, $m$ and $p$ are quark mass and momentum,
similarly $M$ and $P$ for the proton.{\bf \ } With the use of these
distribution functions one can define the function $H$, which in the target
rest frame reads:%
\begin{equation}
H(p_{0})=\sum_{k=1}^{3}e_{k}^{2}\Delta G_{k}(p_{0});\qquad \Delta
G_{k}(p_{0})=G_{k,+1/2}(p_{0})-G_{k,-1/2}(p_{0}),  \label{t4}
\end{equation}%
where $e_{k}$ represent the charges of the proton valence quarks. In the
paper \cite{zav4} I shown, how from the generic function $H$ the spin
structure functions can be obtained. If one assume $Q^{2}\gg 4M^{2}x^{2},$
then:%
\[
g_{1}(x)=\frac{1}{2}\int H(p_{0})\left( m+p_{1}+\frac{p_{1}^{2}}{p_{0}+m}%
\right) \delta \left( \frac{p_{0}+p_{1}}{M}-x\right) \frac{d^{3}p}{p_{0}}%
;\quad x=\frac{Q^{2}}{2M\nu }, 
\]%
\[
g_{2}(x)=-\frac{1}{2}\int H(p_{0})\left( p_{1}+\frac{p_{1}^{2}-p_{T}^{2}/2}{%
p_{0}+m}\right) \delta \left( \frac{p_{0}+p_{1}}{M}-x\right) \frac{d^{3}p}{%
p_{0}}, 
\]%
which implies%
\[
g_{T}(x)\equiv g_{1}(x)+g_{2}(x)=\frac{1}{2}\int H(p_{0})\left( m+\frac{%
p_{T}^{2}/2}{p_{0}+m}\right) \delta \left( \frac{p_{0}+p_{1}}{M}-x\right) 
\frac{d^{3}p}{p_{0}}. 
\]%
Let me remark, that procedure for obtaining the functions $g_{1},g_{2}$ from
the distribution $H$ is rather complex, nevertheless the task is
well-defined and unambiguous. Resulting structure functions are related to a
naive QPM, in which the relativistic kinematics and spheric symmetry (which
follows from the requirement $J=1/2$) are consistently applied. Both these
requirements are very important.

\section{Sum rules}

One can observe, that the functions above have the same general form%
\begin{equation}
\int H(p_{0})f(p_{0},p_{1},p_{T})\delta \left( \frac{p_{0}+p_{1}}{M}%
-x\right) d^{3}p  \label{s1}
\end{equation}%
and differ only in kinematic term $\ f$. This integral, due to spheric
symmetry and presence of the $\delta -$function term, can be expressed as a
combination of the momenta: 
\begin{equation}
V_{n}(x)=\int H(p_{0})\left( \frac{p_{0}}{M}\right) ^{n}\delta \left( \frac{%
p_{0}+p_{1}}{M}-x\right) d^{3}p.  \label{s2}
\end{equation}%
One can prove \cite{zav5}, that these functions satisfy%
\[
\frac{V_{j}^{\prime }(x)}{V_{k}^{\prime }(x)}=\left( \frac{x}{2}+\frac{%
x_{0}^{2}}{2x}\right) ^{j-k};\qquad x_{0}=\frac{m}{M}. 
\]%
This relation then gives possibility to obtain integral relations between
different functions having form (\ref{s2}) or (\ref{s1}), in particular for $%
g_{1}(x)$ and $g_{2}(x)$ one gets:

\[
g_{2}(x)=-\frac{x-x_{0}}{x}g_{1}(x)+\frac{x\left( x+2x_{0}\right) }{\left(
x+x_{0}\right) ^{2}}\int_{x}^{1}\frac{y^{2}-x_{0}^{2}}{y^{3}}g_{1}(y)dy, 
\]%
\[
g_{1}(x)=-\frac{x}{x-x_{0}}g_{2}(x)-\frac{x+2x_{0}}{x^{2}-x_{0}^{2}}%
\int_{x}^{1}g_{2}(y)dy 
\]%
and for limiting case $m\rightarrow 0$:%
\[
g_{2}(x)=-g_{1}(x)+\int_{x}^{1}\frac{g_{1}(y)}{y}dy, 
\]%
\[
g_{1}(x)=-g_{2}(x)-\frac{1}{x}\int_{x}^{1}g_{2}(y)dy. 
\]%
Obviously, the first relation is the known expression for Wanzura - Wilczek
twist-2 term for $g_{2}$ approximation \cite{wawi}.

Further, if one define 
\[
\left\langle x^{\alpha }\right\rangle =\int_{0}^{1}x^{\alpha }V_{0}(x)dx, 
\]%
then one can prove that 
\[
\int_{0}^{1}x^{\alpha }\left[ g_{1}(x)+g_{2}(x)\right] dx=\left\langle
x^{\alpha }\right\rangle \frac{\alpha +1}{\left( \alpha +2\right) \left(
\alpha +3\right) }, 
\]%
\[
\int_{0}^{1}x^{\alpha }g_{2}(x)dx=-\left\langle x^{\alpha }\right\rangle 
\frac{\alpha \left( \alpha +1\right) }{\left( \alpha +2\right) \left( \alpha
+3\right) } 
\]%
for {\it any} $\alpha $, for which the integrals exist. Apparently these
relations imply%
\[
\int_{0}^{1}x^{\alpha }\left[ \frac{\alpha }{\alpha +1}g_{1}(x)+g_{2}(x)%
\right] dx=0, 
\]%
which for $\alpha =2,4,6,...$ corresponds to the Wanzura - Wilczek sum rules %
\cite{wawi}. Other special cases correspond to the Burkhardt - Cottingham ($%
\alpha =0$) \ \cite{buco} and the Efremov - Leader - Teryaev (ELT, $\alpha
=1 $) \cite{elt} sum rules. Let me point out, that all these rules here were
obtained only on the basis of covariant kinematics and requirement of
rotational symmetry.

\section{Valence quarks}

Now I shall try to apply the suggested approach to the description of the
real proton. For simplicity I assume:

1) Spin contribution from the sea of quark-antiquark pairs and gluons can be
neglected, so the proton spin is generated only by the valence quarks.

2) In accordance with the non-relativistic {\it SU(6)} approach, the spin
contribution of individual valence terms is given by fractions:%
\begin{equation}
s_{u}=4/3,\qquad s_{d}=-1/3.  \label{t69}
\end{equation}%
If the symbols $h_{u}$ and $h_{d}$ denote momenta distributions of the
valence quarks in the proton rest frame, which are normalized as%
\begin{equation}
\frac{1}{2}\int h_{u}(p_{0})d^{3}p=\int h_{d}(p_{0})d^{3}p=1,  \label{t70}
\end{equation}%
then the generic distribution (\ref{t4}) reads%
\begin{equation}
H(p_{0})=\sum e_{j}^{2}\Delta h_{j}(p_{0})=\left( \frac{2}{3}\right) ^{2}%
\frac{2}{3}h_{u}(p_{0})-\left( \frac{1}{3}\right) ^{2}\frac{1}{3}%
h_{d}(p_{0}).  \label{t71}
\end{equation}

\begin{figure}
\begin{center}
\epsfig{file=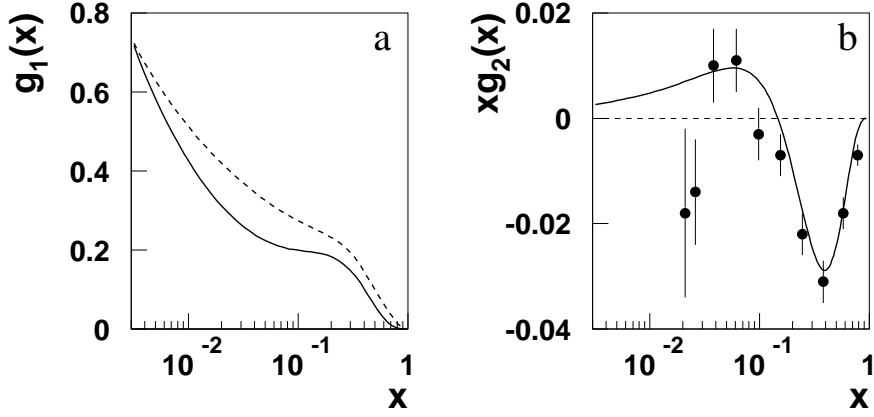, height=7cm}
\end{center}
\caption{Proton spin structure functions. My calculation, which is
represented by the full lines, is compared with the experimental data:
dashed line ($g_{1}$) and full circles ($g_{2}$), see text.}
\label{gps1}
\end{figure}

In the paper \cite{zav1}, using a similar approach, I studied also the
unpolarized structure functions. Structure function $F_{2}$ can be expressed
as%
\begin{equation}
F_{2}(x)=x^{2}\int G(p_{0})\frac{M}{p_{0}}\delta \left( \frac{p_{0}+p_{1}}{M}%
-x\right) d^{3}p;\qquad G(p_{0})=\sum_{q}e_{q}^{2}h_{q}(p_{0}).  \label{t72}
\end{equation}%
On the other hand, for proton valence quarks one can write%
\begin{equation}
F_{2}(x)=\frac{4}{9}xu_{V}(x)+\frac{1}{9}xd_{V}(x),  \label{t73}
\end{equation}%
so combination of the last two relations gives:%
\begin{equation}
q_{V}(x)=x\int h_{q}(p_{0})\frac{M}{p_{0}}\delta \left( \frac{p_{0}+p_{1}}{M}%
-x\right) d^{3}p;\qquad q=u,d.  \label{t74}
\end{equation}%
Since this is again the integral having the structure (\ref{s1}), one can
apply the technique of integral transforms and (instead of relation between $%
g_{1}$ and $g_{2}$) obtain the relations between $g_{j}^{q}$ and $q_{V}$.
For $m\rightarrow 0$ these relations read: 
\[
g_{1}^{q}(x)=\frac{1}{2}\left[ \allowbreak q_{V}(x)-2x^{2}\int_{x}^{1}\frac{%
q_{V}(y)}{y^{3}}dy\right] ,
\]%
\[
g_{2}^{q}(x)=\frac{1}{2}\left[ -\allowbreak \allowbreak
q_{V}(x)+3x^{2}\int_{x}^{1}\frac{q_{V}(y)}{y^{3}}dy\right] .
\]%
Now, taking quark charges and corresponding $SU(6)$ factors as in Eq. (\ref%
{t71}), one can directly calculate $g_{1},g_{2}$ only using the input on the
valence quark distribution $q_{V}=u_{V},d_{V}$. In Fig. \ref{gps1} the
results of $g_{1}$ and $g_{2}$ calculation are shown. 

Experimental data on $%
g_{1}$ are represented by the new parameterization of the world data  \cite%
{e155g1} and the $g_{2}$ points are data of  the E155 Collaboration \cite%
{e155g2}. More detailed discussion of these figures is done in \cite{zav5},
in this talk I want concentrate on the discussion and explanation, why
intrinsic quark motion substantially reduces the first moment of the spin
function $g_{1}$. In \cite{zav4} it is shown, that

\begin{equation}
\Gamma _{1}\equiv \int g_{1}(x)dx=\frac{1}{2}\int H(p_{0})\left( \frac{1}{3}+%
\frac{2m}{3p_{0}}\right) d^{3}p,  \label{v1}
\end{equation}%
which, in the $SU(6)$ approach gives%
\[
\frac{5}{18}\geq \Gamma _{1}\geq \frac{5}{54}, 
\]%
where left limit is valid for the static ($p_{0}\rightarrow m$) and right
one for massless quarks ($m\rightarrow 0$). In other words, it means: 
\[
more\ intrinsic\ motion\Leftrightarrow less\ spin 
\]%
This is a mathematical result, but how to understand it from the point of
view of physics?

First, forget structure functions for a while and calculate completely
another task.\ Let me remind general rules concerning angular momentum in
quantum mechanics:

1) Angular momentum consist of orbital and spin part: {\bf j=l+s}

2) In the relativistic case {\bf l} and {\bf s} are not conserved
separately, only total angular momentum {\bf j} is conserved. So, one can
have pure states of $j(j^{2},j_{z})$ only, which are for fermions with $s=1/2
$ represented by the relativistic spheric waves, see e.g. \cite{lali}:

\[
\psi _{jlj_{z}}\left( {\bf p}\right) =\frac{1}{\sqrt{2p_{0}}}\left( 
\begin{array}{c}
i^{-l}\sqrt{p_{0}+m}\Omega _{jlj_{z}}\left( \frac{{\bf p}}{p}\right) \\ 
i^{-l^{\prime }}\sqrt{p_{0}-m}\Omega _{jl^{\prime }j_{z}}\left( \frac{{\bf p}%
}{p}\right)%
\end{array}%
\right) ;\qquad j=l\pm \frac{1}{2},\qquad l^{\prime }=2j-l, 
\]%
\begin{eqnarray*}
\Omega _{l+1/2,l,j_{z}}\left( \frac{{\bf p}}{p}\right) &=&\left( 
\begin{array}{c}
\sqrt{\frac{j+j_{z}}{2j}}Y_{l,j_{z}-1/2} \\ 
\sqrt{\frac{j-j_{z}}{2j}}Y_{l,j_{z}+1/2}%
\end{array}%
\right) ,\qquad \\
\Omega _{l^{\prime }-1/2,l^{\prime },j_{z}}\left( \frac{{\bf p}}{p}\right)
&=&\left( 
\begin{array}{c}
-\sqrt{\frac{j-j_{z}+1}{2j+2}}Y_{l^{\prime },j_{z}-1/2} \\ 
\sqrt{\frac{j+j_{z}+1}{2j+2}}Y_{l^{\prime },j_{z}+1/2}%
\end{array}%
\right) .
\end{eqnarray*}%
This wavefunction is simplified for the state with total angular momentum
(spin) equal 1/2:

\[
j=j_{z}=\frac{1}{2},\qquad l=0\qquad \Rightarrow \qquad l^{\prime }=1, 
\]%
\[
Y_{00}=\frac{1}{\sqrt{4\pi }},\qquad Y_{10}=i\sqrt{\frac{3}{4\pi }}\cos
\theta ,\qquad Y_{11}=-i\sqrt{\frac{3}{8\pi }}\sin \theta \exp \left(
i\varphi \right) , 
\]%
which gives%
\[
\psi _{jlm}\left( {\bf p}\right) =\frac{1}{\sqrt{8\pi p_{0}}}\left( 
\begin{array}{c}
\sqrt{p_{0}+m}\left( 
\begin{array}{c}
1 \\ 
0%
\end{array}%
\right) \\ 
-\sqrt{p_{0}-m}\left( 
\begin{array}{c}
\cos \theta \\ 
\sin \theta \exp \left( i\varphi \right)%
\end{array}%
\right)%
\end{array}%
\right) . 
\]%
Let me remark, that $j=1/2$ is minimum angular momentum for particle with $%
s=1/2.$ Now, one can easily calculate the average contribution of the spin
operator to the total angular momentum:

\[
\Sigma _{3}=\frac{1}{2}\left( 
\begin{array}{cc}
\sigma _{3} & \cdot \\ 
\cdot & \sigma _{3}%
\end{array}%
\right) \Rightarrow 
\]%
\[
\psi _{jlm}^{\dagger }\left( {\bf p}\right) \Sigma _{3}\psi _{jlm}\left( 
{\bf p}\right) =\frac{1}{16\pi p_{0}}\left[ \left( p_{0}+m\right) +\left(
p_{0}-m\right) \left( \cos ^{2}\theta -\sin ^{2}\theta \right) \right] 
\]%
If $a_{p}$ is the probability amplitude of the state $\psi _{jlm}$, then 
\begin{equation}
\left\langle \Sigma _{3}\right\rangle =\int a_{p}^{\star }a_{p}\psi
_{jlm}^{\dagger }\left( {\bf p}\right) \Sigma _{3}\psi _{jlm}\left( {\bf p}%
\right) d^{3}p=\frac{1}{2}\int a_{p}^{\star }a_{p}\left( \frac{1}{3}+\frac{2m%
}{3p_{0}}\right) p^{2}dp,  \label{v2}
\end{equation}%
which means, that:

{\it i)} For the fermion at rest ($p_{0}=m$) we have $j=s=1/2,$ which is
quite comprehensible, since without kinetic energy no orbital momentum can
be generated.

{\it ii)}\ For the state in which $p_{0}\geq m$, we have in general: 
\[
\frac{1}{3}\leq \frac{\left\langle s\right\rangle }{j}\leq 1. 
\]%
where left limit is valid for the energetic fermion, $p_{0}\gg m$. In other
words, in the states $\psi _{jlm}$ with $p_{0}>m$ part of the total angular
momentum $j=1/2$ is {\it necessarily }created by orbital momentum. This is a
simple consequence of quantum mechanics.

Now, one can compare integrals (\ref{v1}) and (\ref{v2}). Since both
integrals involve the same kinematic term, the interpretation of dependence
on ratio $m/p_{0}$ in (\ref{v2}) is valid also for (\ref{v1}). Otherwise,
the comparison is a rigorous illustration of the statement, that $\Gamma
_{1} $ measures contributions from quark spins (and not their total angular
momenta).

In which point the present approach differ from standard QPM? Standard
approach is closely connected with the preferred reference frame - infinite
momentum frame. The basic relations like 
\[
g_{1}(x)=\frac{1}{2}\sum e_{j}^{2}\Delta q_{j}(x),\qquad F_{2}(x)=x\sum
e_{i}^{2}q_{i}(x) 
\]%
are derived with the use of approximation%
\[
p_{\alpha }=xP_{\alpha }. 
\]%
In the covariant formulation this relation is equivalent to the assumption,
that the quarks are static. In the presented covariant approach quarks are
not static, so this approximation cannot be used. As a result, different
relations between the distribution and structure functions and also
different behavior of $\Gamma _{1}$ are obtained.

\section{Summary}

I have studied spin functions in system of quasifree fermions having
fixed effective mass $x_{0}=m/M$ and total spin $J=1/2$ - representing a
covariant version of naive QPM. The main results are:

1) Spin functions $g_{1}$ and $g_{2}$ depend on intrinsic motion. In
particular, the momenta $\Gamma _{1}$ corresponding to the static (massive)
fermions and massless fermions, can differ significantly: $\Gamma _{1}(m\ll
p_{0})/\Gamma _{1}(p_{0}\approx m)=1/3$. It is due to splitting of angular
momentum into spin and orbital part, as soon as intrinsic motion is present.

2) $g_{1}$ and $g_{2}$ are connected by a simple transformation, which is
for $m\rightarrow 0$ identical to Wanzura - Wilczek relation for twist-2
term of the $g_{2}$ approximation. Relations for the $n-th$ momenta of the
structure functions have been obtained, their particular cases are identical
to known sum rules: Wanzura - Wilczek ($n=2,4,6...$), Efremov - Leader -
Teryaev ($n=1$) and Burkhardt - Cottingham ($n=0$).

3) Model has been applied to the proton spin structure, assuming proton spin
is generated only by spins and orbital momenta of the valence quarks with 
{\it SU(6)} symmetry and for quark effective mass $m\rightarrow 0$. As an
input I used \ known parameterization of the valence terms, then without any
other free parameter, the functions $g_{1},$ $g_{2}$ were obtained.
Comparison with the proton data demonstrates a good agreement.

4) Comparison with the corresponding relations for the structure functions
following from the usual naive QPM was done. Both the approaches are
equivalent for the static quarks. Differences for quarks with internal
motion inside the proton are result of the conflict with the assumption $%
p_{\alpha }=xP_{\alpha }$, which is crucial for derivation of the relations
in the standard QPM.

\end{document}